\documentclass{article}

\usepackage{spconf,amsmath,epsfig,amsfonts}
\usepackage{psfrag}
\usepackage{amsmath}
\usepackage{amssymb}
\usepackage{amsfonts}
\usepackage{color}
\usepackage{tabularx}
\usepackage{subfigure}

\newcolumntype{C}[1]{>{\centering\arraybackslash}p{#1}}

\newcount\hour \newcount\minute
\hour=\time \divide \hour by 60
\minute=\time
\advance \minute by \count99
\newcommand{\daytime}{%
  \ifnum\hour=0 00\else\ifnum\hour<10 0\fi\number\hour\fi:%
  \ifnum\minute<10 0\fi\number\minute%

}

\definecolor{MyGrey}{rgb}{0.5,0.5,0.5}

\usepackage{subfigure}
\renewcommand{\thesubfigure}{\thefigure.\arabic{subfigure}}
\makeatletter
\renewcommand{\p@subfigure}{}
\renewcommand{\@thesubfigure}{{\bf Fig. \thesubfigure}.\ }
\makeatother

\def\ve#1{{\mathchoice{\mbox{\boldmath$\displaystyle #1$}}%
		      {\mbox{\boldmath$\textstyle #1$}}%
		      {\mbox{\boldmath$\scriptstyle #1$}}%
		      {\mbox{\boldmath$\scriptscriptstyle #1$}}}} 

\def\diag{\mathrm{diag}}
\newcommand{\argmax}{\mathop{\mathrm{argmax}}}
\def\PSNR{\mathrm{ PSNR}}
\def\punit{\, \mathrm}

\title{Fast orthogonality deficiency compensation for improved frequency selective image extrapolation}
\name{J\"urgen~Seiler and Andr\'e~Kaup}
\address{Chair of Multimedia Communications and Signal Processing, \\University of Erlangen-Nuremberg, Cauerstr. 7, 91058 Erlangen, Germany\\
{\{seiler, kaup\}@LNT.de}}

\begin{document}
\topmargin=0mm
\ninept
\maketitle


\begin{abstract} \label{abstract}
The purpose of this paper is to introduce a very efficient algorithm for signal extrapolation. It can widely be used in many applications in image and video communication, e.\ g.\ for concealment of block errors caused  by transmission errors or for prediction in video coding. The signal extrapolation is performed by extending a signal from a limited number of known samples into areas beyond these samples. Therefore a finite set of orthogonal basis functions is used and the known part of the signal is projected onto them. Since the basis functions are not orthogonal regarding the area of the known samples, the projection does not lead to the real portion a basis function has of the signal. The proposed algorithm efficiently copes with this non-orthogonality resulting in very good objective and visual extrapolation results for edges, smooth areas, as well as structured areas. Compared to an existent implementation, this algorithm has a significantly lower computational complexity without any degradation in quality. The processing time can be reduced by a factor larger than $100$.
\end{abstract}




\vspace{-2mm} \section{Introduction} \label{sec:introduction} \vspace{-2mm}

The estimation of data samples from known surrounding samples is an important task in many modern communication applications. Extending a discrete signal from known areas into areas where no amplitude information is accessible is usually called signal extra\-polation. In image and video communication a common application for signal extrapolation is concealment of block losses by estimating lost areas from correctly received adjacent areas. Signal extrapolation could as well be used for signal prediction whereas data samples are estimated based on already known samples. So only the prediction error between the original samples and the estimated samples has to be transmitted.

In \cite{Seiler2007} we presented the orthogonality deficiency compensated frequency selective extrapolation (OFSE), an efficient algorithm for signal extrapolation based on the frequency selective extrapolation (FSE) proposed in \cite{Meisinger2004b}. We showed that this algorithm provides very good extrapolation results for concealment of block errors. The extrapolation results were compared to the ones from existing concealment algorithms such as the maximally smooth image recovery algorithm by Wang et al. \cite{Wang1993}, the projections onto convex sets (POCS) algorithm proposed by Sun and Kwok \cite{Sun1995}, the DCT-based interpolation algorithm by \mbox{Alkachouh} and Bellanger \cite{Alkachouh2000} and the sequential error-concealment algorithm by Li and Orchard \cite{Li2002}. Even if we had been able to gain a large increase in $\PSNR$ and to reduce visual artifacts, the algorithm suffers from its high computational complexity and processing time. In this paper, we will present the fast orthogonality deficiency compensated frequency selective extrapolation (FOFSE), a modification of the algorithm from \cite{Seiler2007} that needs far less operations whereas the extrapolation quality is still on the same high level.

In the following, we start with a short review of the OFSE algorithm \cite{Seiler2007} in order to identify two computational very expensive steps. Subsequently, we propose a method to reduce the operations in these parts and compare the extrapolation results in terms of $\PSNR$ and processing time with the original OFSE algorithm and the algorithms mentioned above. The algorithm is carried out only for two-dimensional data sets, but by making use of \cite{Meisinger2007} it could be adapted to three-dimensional sets as well.


\vspace{-2mm} \section{Signal extrapolation} \label{sec:extrapolation} \vspace{-2mm}

\begin{figure}
	\psfrag{m}[c][c][0.8]{$m$}
	\psfrag{n}[c][c][0.8]{$n$}
	\psfrag{M}[c][c][0.8]{$M$}
	\psfrag{N}[c][c][0.8]{$N$}
	\psfrag{L}[l][l][0.8]{$\mathcal{L} = \mathcal{A} \cup \mathcal{B}$}
	\psfrag{A}[l][l][0.8]{$\mathcal{A}$}
	\psfrag{B}[l][l][0.8]{$\mathcal{B}$}
	\centering
	\includegraphics[width=0.10\textwidth]{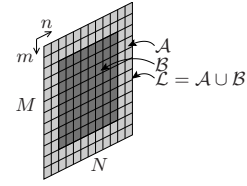}
	\vspace{-0.2cm}\caption{Data area $\mathcal{L}$ used for two-dimensional extrapolation consisting of the missing area to be estimated $\mathcal{B}$ and the known surrounding support area $\mathcal{A}$.}
	\label{fig:region2d}
\end{figure}

Fig.\ \ref{fig:region2d} shows a possible two-dimensional data set, depicted by the two spatial coordinates $m$ and $n$. In area $\mathcal{B}$, called missing area, the data samples of unknown magnitude are subsumed. The idea of signal extrapolation is to estimate these samples by means of the data samples with known magnitude. These samples are subsumed in area $\mathcal{A}$, called the support area. Both areas $\mathcal{A}$ and $\mathcal{B}$ together form area $\mathcal{L}$ containing all data samples being involved in the extrapolation process.

We regard the discrete function $f\left[m,n\right]$ that is defined over the whole area $\mathcal{L}$. The magnitudes from function $f\left[m,n\right]$ are only accessible over the support area $\mathcal{A}$ and we aim to get the magnitudes over area $\mathcal{B}$ by generating a parametric model $g\left[m,n\right]$ that is also defined over $\mathcal{L}$. The parametric model is generated by a weighted linear combination of mutually orthogonal two-dimensional basis functions $\varphi_k\left[m,n\right]$.
\begin{equation}
g\left[m,n\right] = \sum_{\forall k \in \mathcal{K}} c_k \varphi_k \left[m,n\right] 
\end{equation}
Thereby the set $\mathcal{K}$ covers the indices of all used basis functions and the weighting factors $c_k$ are denoted as expansion coefficients. The algorithm aims to generate $g\left[m,n\right]$ in a way that it becomes a good approximation of $f\left[m,n\right]$ in $\mathcal{A}$. As $g\left[m,n\right]$ is defined over whole $\mathcal{L}$ we get a signal continuation into area $\mathcal{B}$. The parametric model is generated in an iterative approach whereas in every iteration step one basis function with its corresponding expansion coefficient is added
\begin{equation}
g^{\left(\nu \right)} \left[m,n\right] = g^{\left(\nu-1 \right)} \left[m,n\right] + \hat{c}_u^{\left(\nu \right)} \cdot \varphi_u \left[m,n\right].
\end{equation}
$u$ is the index of the basis function that was chosen for this iteration step and $\hat{c}_u^{\left(\nu \right)}$ is the estimate for the expansion coefficient. $g^{\left(\nu \right)} \left[m,n\right]$ denotes the parametric model in the $\nu$-th iteration step. Initially, the model $g^{\left(0 \right)} \left[m,n\right]$ is set to zero

The arising residual approximation error $r^{\left(\nu \right)} \left[m,n\right]$ in the $\nu$-th iteration step between $f\left[m,n\right]$ and $g^{\left(\nu \right)} \left[m,n\right]$ is
\begin{equation}
r^{\left(\nu \right)} \left[m,n\right] = \left\{ \begin{array}{ll}\hspace{-2mm} r^{\left(\nu - 1  \right)} \left[m,n\right] - \hat{c}_u^{\left(\nu \right)} \cdot \varphi_u \left[m,n\right] &\hspace{-2mm},\forall \left(m,n\right) \in \mathcal{A} \\ \hspace{-2mm}0 &\hspace{-2mm}, \forall \left(m,n\right) \in \mathcal{B} \end{array} \right.
\end{equation}

In every iteration step a weighted projection of $r^{\left(\nu \right)} \left[m,n\right]$ onto each basis function is performed. Thereby the weighting function
\begin{equation}
w \left[m,n\right] = \left\{ \begin{array}{ll} \rho\left[m,n\right] &,\ \forall \left(m,n\right) \in \mathcal{A} \\ 0 &,\ \forall \left(m,n\right) \in \mathcal{B} \end{array} \right.
\end{equation}
is used to control the influence a sample has on the extrapolation process depending on its location. On the one hand $w\left[m,n\right]$ is used to mask area $\mathcal{B}$, on the other hand it performs the actual weighting by $\rho\left[m,n\right]$ which can be chosen arbitrarily. A good choice for $\rho\left[m,n\right]$ is given in (\ref{eq:isotropic_model}). The weighted projection onto the $k$-th basis function yields the projection coefficient $p_k^{\left(\nu \right)}$.
\begin{equation}
\label{eq:proj_coeff}
p_k^{\left(\nu \right)} = \frac{\displaystyle \sum_{\left(m,n\right) \in \mathcal{L}} r^{\left(\nu-1\right)} \left[m,n\right] \cdot \varphi_k \left[m,n\right] \cdot w \left[m,n\right]}{\displaystyle \sum_{\left(m,n\right) \in \mathcal{L}} w \left[m,n\right] \cdot \varphi_k^2 \left[m,n\right] } .
\end{equation}
Hereby the numerator is the weighted scalar product between the approximation error and the $k$-th basis function. The numerator further is normalized by the weighted scalar product between the selected basis function and itself.

According to \cite{Meisinger2004b} and \cite{Seiler2007} the basis function to be added to the parametric model in an iteration step is the one that minimizes the distance between the error signal and the weighted projection onto the basis function. This results in index $u$ of the basis function determined by
\begin{equation}
u = \argmax_{k=0,\ldots,\left|\mathcal{L}\right|-1} \left( p_k^{\left(\nu\right)^2} \cdot \sum_{\left(m,n\right) \in \mathcal{L}} w\left[m,n\right] \cdot \varphi^2_k \left[m,n\right] \right).
\end{equation}
Obviously, this criterion for chosing the basis function is computationally very expensive as all projection coefficients have to be computed and have to be compared. Hence, for the FOFSE we propose to use the basis function that forms the biggest absolut portion of the weighted residual error 
\begin{equation}
r_{\mathrm{w}}^{\left(\nu\right)}\left[m,n\right]=r^{\left(\nu\right)}\left[m,n\right]\cdot w\left[m,n\right].
\end{equation}
Therefore we regard the operator $\mathcal{T}$ that performs a decomposition of a two-dimensional spatial signal into the used set of basis functions. The operator returns a vector of scalars that quantifies the portion each basis function has of the signal.
\begin{equation}
\ve{R}_{\mathrm{w}}^{\left(\nu\right)} = \mathcal{T} \left\{ r_{\mathrm{w}}^{\left(\nu\right)}\left[m,n\right] \right\}
\end{equation}
The index $u$ of the basis function to use is then determined by
\begin{equation}
u = \argmax_{k=0,\ldots,\left|\mathcal{L}\right|-1} \left| \ve{R}_{\mathrm{w}}^{\left(\nu\right)}\left[k\right] \right |
\end{equation}
Depending on the used set of basis functions, efficient algorithms for decomposing a signal into the basis functions exist and thus $r_{\mathrm{w}}^{\left(\nu\right)}\left[m,n\right]$ only has to be transformed into the domain of the basis functions and a maximum has to be found. With several sets of bases, such as the functions of the two-dimensional DFT (compare \cite{Meisinger2004b}) or the two-dimensional DCT, a complete formulation of the algorithm in the transform domain is possible as well, and $r_{\mathrm{w}}^{\left(\nu\right)}\left[m,n\right]$ does not have be transformed in every iteration step. 

In the next step, the estimate $\hat{c}_u^{\left(\nu\right)}$ for the just chosen basis function has to be determined. Unfortunately, the basis functions are not mutually orthogonal when evaluated with respect to the support area $\mathcal{A}$ in combination with the weighting function. Due to this fact, the projection onto a basis function does not only lead to the portion this basis function has of the approximation error but in addition, portions of other basis functions are incorporated as well. For determining $\hat{c}_u^{\left(\nu\right)}$, the estimate of the real portion a basis function has of the approximation error signal, this orthogonality deficiency has to be compensated. 

The orthogonality deficiency compensation proposed in \cite{Seiler2007} obtains the compensation by calculating all projection coefficients $p_k^{\left(\nu\right)},$ $k=0,\ldots, \left| \mathcal{L}\right|-1$ and determines $\hat{c}_u^{\left(\nu\right)}$ according to
\begin{equation}
\label{eq:elaborate_compensation}
\hat{c}^{\left(\nu\right)}_u = \frac{p^{\left(\nu\right)}_u}{\displaystyle \sum_{l=0,\ldots,\left|\mathcal{L}\right|-1} \frac{p^{\left(\nu\right)}_l}{p^{\left(\nu\right)}_u} \cdot \left( \hat{\ve{K}} \right)_{u,l}}
\end{equation}
Here $\left( \hat{\ve{K}} \right)_{u,l}$ denotes the $l$-column in the $u$-th line of matrix $\hat{\ve{K}}$ with matrix $\hat{\ve{K}}$ emanating from 
\begin{equation}
\hat{\ve{K}} = \left(\diag \left( \diag \left( \ve{K} \right) \right) \right)^{-1} \cdot \ve{K}.
\end{equation}
The square matrix $\ve{K}$ is the matrix of the weighted scalar products of all basis functions.
\begin{equation}
\ve{K} = \left( \begin{array}{ccc}
		\displaystyle \sum_{\left(m,n\right) \in \mathcal{L}} \tilde{w}  \tilde{\varphi}_0  \tilde{\varphi}_0  & \cdots & \displaystyle \sum_{\left(m,n\right) \in \mathcal{L}} \tilde{w}  \tilde{\varphi}_{\left|\mathcal{L}\right|-1}   \tilde{\varphi}_0  \\
		\vdots & \ddots & \vdots \\
		\displaystyle \sum_{\left(m,n\right) \in \mathcal{L}} \tilde{w}  \tilde{\varphi}_0  \tilde{\varphi}_{\left|\mathcal{L}\right|-1}  & \cdots & \displaystyle \sum_{\left(m,n\right) \in \mathcal{L}} \tilde{w}  \tilde{\varphi}_{\left|\mathcal{L}\right|-1}   \tilde{\varphi}_{\left|\mathcal{L}\right|-1} 
                \end{array} \right)
\end{equation}
In this equation two abbreviations are used, $\tilde{w} = w\left[m,n\right]$ and $\tilde{\varphi_i}= \varphi_i\left[m,n\right]$.

Although OFSE provides very good estimates for the expansion coefficients, it is computationally very expensive. The two main reasons are the fact that all possible projection coefficients have to be calculated and the circumstance that many operations are necessary to generate $\hat{\ve{K}}$ and $\hat{c}_u^{\left(\nu\right)}$. 

\begin{figure}
\centering
\psfrag{s01}[t][t]{\color[rgb]{0,0,0}\setlength{\tabcolsep}{0pt}\begin{tabular}{c}Orthogonality compensation factor\end{tabular}}%
\psfrag{s02}[b][b]{\color[rgb]{0,0,0}\setlength{\tabcolsep}{0pt}\begin{tabular}{c}Relative occurence\end{tabular}}%
\psfrag{s04}[b][b]{\color[rgb]{0,0,0}\setlength{\tabcolsep}{0pt}}
\psfrag{s06}[][]{\color[rgb]{0,0,0}\setlength{\tabcolsep}{0pt}\begin{tabular}{c} \end{tabular}}%
\psfrag{s07}[][]{\color[rgb]{0,0,0}\setlength{\tabcolsep}{0pt}\begin{tabular}{c} \end{tabular}}%
\psfrag{s08}[l][l][0.75]{\color[rgb]{0,0,0}``Peppers''}%
\psfrag{s21}[l][l][0.75]{\color[rgb]{0,0,0}``Baboon''}%
\psfrag{s22}[l][l][0.75]{\color[rgb]{0,0,0}``Lena''}%
\psfrag{s23}[l][l][0.75]{\color[rgb]{0,0,0}``Peppers''}%
\psfrag{x12}[t][t][0.75]{$0$}%
\psfrag{x13}[t][t][0.75]{$0.1$}%
\psfrag{x14}[t][t][0.75]{$0.2$}%
\psfrag{x15}[t][t][0.75]{$0.3$}%
\psfrag{x16}[t][t][0.75]{$0.4$}%
\psfrag{x17}[t][t][0.75]{$0.5$}%
\psfrag{x18}[t][t][0.75]{$0.6$}%
\psfrag{x19}[t][t][0.75]{$0.7$}%
\psfrag{x20}[t][t][0.75]{$0.8$}%
\psfrag{x21}[t][t][0.75]{$0.9$}%
\psfrag{x22}[t][t][0.75]{$1$}%
\psfrag{v12}[r][r][0.75]{$0$}%
\psfrag{v13}[r][r][0.75]{$0.1$}%
\psfrag{v14}[r][r][0.75]{$0.2$}%
\psfrag{v15}[r][r][0.75]{$0.3$}%
\psfrag{v16}[r][r][0.75]{$0.4$}%
\psfrag{v17}[r][r][0.75]{$0.5$}%
\includegraphics[width=0.35\textwidth]{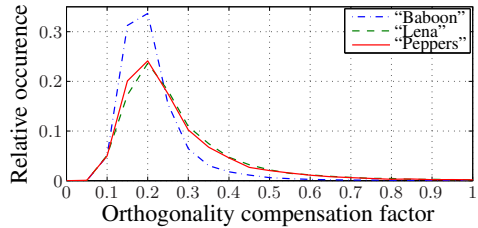}
\vspace{-0.2cm}\caption{Distribution of the orthogonality deficiency compensation factor.}
	\label{fig:oc_distribution}
\end{figure}

Examining (\ref{eq:elaborate_compensation}) for different scenarios, we recognized that the compensation factor 
\begin{equation}
\gamma^{\left(\nu\right)}_u = \frac{1}{\displaystyle \sum_{l=0,\ldots,\left|\mathcal{L}\right|-1} \frac{p^{\left(\nu\right)}_l}{p^{\left(\nu\right)}_u} \cdot \left( \hat{\ve{K}} \right)_{u,l}}
\end{equation}
is from a very small range of values. The center of this range strongly depends on the extrapolation scenario. If support area $\mathcal{A}$ is much larger than loss area $\mathcal{B}$ the basis functions are still close to orthogonality, resulting in compensation factors close to $1$. With decreasing size of $\mathcal{A}$ the values of $\gamma^{\left(\nu\right)}_u$ tend towards $0$. As an example \mbox{Fig. \ref{fig:oc_distribution}} shows the occuring orthogonality deficiency compensation factors for extrapolation of $16\times16$ pixels sized blocks framed by a support area of $16$ pixels width. $200$ iterations are performed and $81$ block losses are considered. Test images are ``Baboon'', ``Lena'' and ``Peppers''. The basis functions used, are the ones from the two-dimensional discrete Fourier transform and a FFT of size $64\times64$ is used. Apparently, in most cases a compensation factor about $0.2$ is calculated.

Therefore we propose to apply a constant compensation factor $\gamma$ between $0$ and $1$, independent from the considered basis function and the iteration step. So the compensation in (\ref{eq:elaborate_compensation})  can be simplified and we obtain
\begin{equation}
\hat{c}^{\left(\nu\right)}_u = \gamma \cdot p^{\left(\nu\right)}_u
\end{equation}
As we will show later, the exact choice for $\gamma$ is not critical and by adjusting $\gamma$, the extrapolation can be tuned between complexity and quality.


\vspace{-2mm} \section{Complexity valuation}\label{sec:valuation} \vspace{-2mm}

Subsequently a valuation of the number of operations for the fast orthogonality deficiency compensated frequency selective extrapolation (FOFSE) is compared to the number of operations for the original orthogonality deficiency compensated extrapolation (OFSE) \cite{Seiler2007}. As basis functions we use the functions of the two-dimensional Fourier transform as there exists an efficient implementation in the transform domain \cite{Meisinger2004b}. We regard a block of $M\times N$ samples and we will use an FFT of size $T\times T$ for the transform into the frequency domain. The number of iterations is indicated by $I$. We assume hypothetical runtime optimized realizations, meaning that everything that can be computed in advance is computed in advance. These hypothetical realizations are extremly memory consuming but the overall numbers of operations are minimal. 

\begin{table}{\hfuzz=\maxdimen \newdimen\hfuzz \centering
\begin{tabular}{|l|c|c|}
\hline & OFSE \cite{Seiler2007} & FOFSE \\
\hline \hline MUL & $MN + I \cdot \left(\frac{49T^2}{2}-16\right)$  & $MN + I \cdot \left(9 T^2-12\right)$ \\ 
\hline MEM & $2MN + I\cdot\left(10T^2+2\right)$& $2MN + I\cdot\left(\frac{7T^2}{2}  +10\right)$  \\ 
\hline ADD & $I\cdot\left(14T^2-16\right)$ & $I\cdot\left(6T^2+5\right)$ \\ 
\hline FUNC & $I\cdot \left(\frac{9T^2}{2}-1\right)$ & $I\cdot \left(\frac{3T^2}{2}+4\right)$ \\ \hline
\end{tabular}
\vspace{-0.2cm}\caption{Number of operations for OFSE and FOFSE.}\vspace{-0.1cm}
\label{tab:operations}}
\end{table}

\begin{figure}
\centering
\psfrag{s01}[t][t]{\color[rgb]{0,0,0}\setlength{\tabcolsep}{0pt}\begin{tabular}{c}Iterations\end{tabular}}%
\psfrag{s02}[b][b]{\color[rgb]{0,0,0}\setlength{\tabcolsep}{0pt}\begin{tabular}{c}Operations / $10^6$\end{tabular}}%
\psfrag{s04}[b][b]{\color[rgb]{0,0,0}\setlength{\tabcolsep}{0pt}}
\psfrag{s06}[][]{\color[rgb]{0,0,0}\setlength{\tabcolsep}{0pt}\begin{tabular}{c} \end{tabular}}%
\psfrag{s07}[][]{\color[rgb]{0,0,0}\setlength{\tabcolsep}{0pt}\begin{tabular}{c} \end{tabular}}%
\psfrag{s08}[l][l][0.75]{\color[rgb]{0,0,0}Function calls (OFSE)}%
\psfrag{s21}[l][l][0.75]{\color[rgb]{0,0,0}FOFSE: MUL}%
\psfrag{s22}[l][l][0.75]{\color[rgb]{0,0,0}\hphantom{FOFSE:} MEM}%
\psfrag{s23}[l][l][0.75]{\color[rgb]{0,0,0}\hphantom{FOFSE:} ADD}%
\psfrag{s24}[l][l][0.75]{\color[rgb]{0,0,0}\hphantom{FOFSE:} FUNC}%
\psfrag{s25}[l][l][0.75]{\color[rgb]{0,0,0}OFSE:\hphantom{F} MUL}%
\psfrag{s26}[l][l][0.75]{\color[rgb]{0,0,0}\hphantom{OFSE:}\hphantom{F} MEM}%
\psfrag{s27}[l][l][0.75]{\color[rgb]{0,0,0}\hphantom{OFSE:}\hphantom{F} ADD}%
\psfrag{s28}[l][l][0.75]{\color[rgb]{0,0,0}\hphantom{OFSE:}\hphantom{F} FUNC}%
\psfrag{x12}[t][t][0.75]{$0$}%
\psfrag{x13}[t][t][0.75]{$5$}%
\psfrag{x14}[t][t][0.75]{$10$}%
\psfrag{x15}[t][t][0.75]{$15$}%
\psfrag{x16}[t][t][0.75]{$20$}%
\psfrag{x17}[t][t][0.75]{$25$}%
\psfrag{x18}[t][t][0.75]{$30$}%
\psfrag{x19}[t][t][0.75]{$35$}%
\psfrag{x20}[t][t][0.75]{$40$}%
\psfrag{x21}[t][t][0.75]{$45$}%
\psfrag{x22}[t][t][0.75]{$50$}%
\psfrag{v12}[r][r][0.75]{$0$}%
\psfrag{v13}[r][r][0.75]{$1$}%
\psfrag{v14}[r][r][0.75]{$2$}%
\psfrag{v15}[r][r][0.75]{$3$}%
\psfrag{v16}[r][r][0.75]{$4$}%
\psfrag{v17}[r][r][0.75]{$5$}%
\psfrag{v18}[r][r][0.75]{$6$}%
\vspace{-0.4cm}\includegraphics[width=0.35\textwidth]{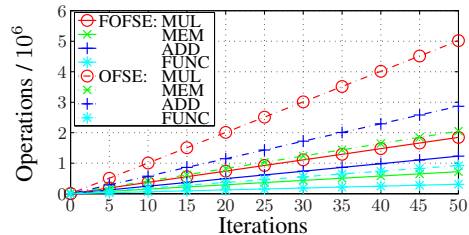}
\vspace{-0.2cm}\caption{Complexity comparison OFSE and FOFSE.}
	\label{fig:ops_over_iter}
\end{figure}

In Table \ref{tab:operations} the required multiplications (MUL), memory accesses (MEM), additions (ADD), and function calls (FUNC) are listed for both algorithms with respect to the number of iterations, the spatial size, and the transform size. Fig. \ref{fig:ops_over_iter} is used to illustrate the necessary number of operations for both approaches with respect to the number of iterations. The extrapolation scenario is the same as described above. In addition to the values mentioned so far, for both approaches $2T$ Fast Fourier Transforms of length $T$ are needed for the transform into the frequency domain and back. Summarized, even for these hypothetical implementations only about one third of the number of operations is needed for the proposed algorithm. As such implementations are not possible on actual processors a realistic implementation is used for the following runtime evaluations. Thereby many of the pre-calculated values have to be replaced with just in time calculated ones. As these calculations are more \mbox{expensive} for OFSE than for FOFSE, the fast approach will even be more than three times faster, as indicated by the hypothetical realizations.

\vspace{-2mm} \section{Results}\label{sec:results} \vspace{-2mm}

\begin{table*}{\hfuzz=\maxdimen \newdimen\hfuzz 
\hspace{1cm}\begin{tabular}{|p{6cm}|C{1.5cm}|C{1.5cm}|C{1.5cm}|C{3.5cm}|} \hline
& {``Lena''} & {``Peppers''} & {``Baboon''} & Processing time per block  \\ \hline \hline
Maximally smooth recovery \cite{Wang1993}& $24.8 \punit{dB}$ & $24.6 \punit{dB}$ & $19.6 \punit{dB}$ & $0.0077 \punit{sec}$ \\ \hline
Spatial domain interpolation \cite{Alkachouh2000} & $22.2 \punit{dB}$ & $23.4 \punit{dB}$ & $16.8 \punit{dB}$ & $0.0036\punit{sec}$ \\ \hline
POCS \cite{Sun1995} & $22.8 \punit{dB}$ & $22.7 \punit{dB}$ & $19.0 \punit{dB}$ & $0.0056\punit{sec}$\\ \hline
Sequential error concealment \cite{Li2002} & $24.7 \punit{dB}$ & $26.9 \punit{dB}$ & $18.7 \punit{dB}$ & $1.78\punit{sec}$ \\ \hline
FSE ($20$ iterations) \cite{Meisinger2004b} & $24.8\punit{dB}$ & $25.2 \punit{dB}$& $18.6 \punit{dB}$& $0.096\punit{sec}$\\ \hline 
OFSE ($200$ iterations) \cite{Seiler2007} & $26.7 \punit{dB}$ & $26.8 \punit{dB}$ & $19.7 \punit{dB}$ & $43.91\punit{sec}$\\ \hline
FOFSE ($200$ iterations) & $26.7 \punit{dB}$ & $26.9 \punit{dB}$ & $19.7 \punit{dB}$ &  $0.25\punit{sec}$\\ \hline
\end{tabular}}
\vspace{-0.2cm}\caption{Maximum achievable $\PSNR$ and processing time per block for different extrapolation algorithms.}\vspace{-0.3cm}
\label{tab:odc_results}
\end{table*}

In the following, the extrapolation results for the fast orthogonality deficiency compensated frequency selective extrapolation (FOFSE) are evaluated by concealing lost blocks in images. Therefore blocks of size $16\times16$ pixels are cut out of the test images ``Baboon'', ``Lena'' and ``Peppers''. These blocks are extrapolated and compared to the original blocks in terms of $\PSNR$. The support area is a frame of $16$ pixels width. Further, the functions of the two-dimensional discrete Fourier transform are used as basis functions since the complete extrapolation algorithm can be performed in the frequency domain \cite{Meisinger2004b}. In addition, these basis functions are well suited for extrapolation as monotone areas, noisy regions and edges can be extrapolated very well. For the transform into the frequency domain a FFT of size $64\times64$ is used. According to \cite{Meisinger2004b, Seiler2007}, the used weighting function is generated by a radial symmetric isotropic model
\begin{equation}
\rho\left[m,n\right] = \hat{\rho}^{\sqrt{\left(m-\frac{M-1}{2}\right)^2 + \left(n-\frac{N-1}{2}\right)^2}}
\label{eq:isotropic_model}
\end{equation}
with the correlation coefficient $\hat{\rho}$ chosen to $0.8$.

\begin{figure}
\centering 
\psfrag{s05}[t][t]{\color[rgb]{0,0,0}\setlength{\tabcolsep}{0pt}\begin{tabular}{c}Iterations\end{tabular}}%
\psfrag{s06}[b][b]{\color[rgb]{0,0,0}\setlength{\tabcolsep}{0pt}\begin{tabular}{c}$\PSNR$ in $\punit{dB}$\end{tabular}}%
\psfrag{s08}[b][b]{\color[rgb]{0,0,0}\setlength{\tabcolsep}{0pt}}
\psfrag{s10}[][]{\color[rgb]{0,0,0}\setlength{\tabcolsep}{0pt}\begin{tabular}{c} \end{tabular}}%
\psfrag{s11}[][]{\color[rgb]{0,0,0}\setlength{\tabcolsep}{0pt}\begin{tabular}{c} \end{tabular}}%
\psfrag{s12}[l][l][0.75]{\color[rgb]{0,0,0}``Baboon'' (FSE)}%
\psfrag{s13}[l][l][0.75]{\color[rgb]{0,0,0}``Lena'':\newlength{\nlena}\settowidth{\nlena}{``Lena'':}\hspace{-\nlena}\hphantom{``Peppers'':} FOFSE, $\gamma=0.2$}%
\psfrag{s14}[l][l][0.75]{\color[rgb]{0,0,0}\hphantom{``Peppers'':} OFSE}%
\psfrag{s15}[l][l][0.75]{\color[rgb]{0,0,0}\hphantom{``Peppers'':} FSE}%
\psfrag{s16}[l][l][0.75]{\color[rgb]{0,0,0}``Peppers'': FOFSE, $\gamma=0.2$}%
\psfrag{s17}[l][l][0.75]{\color[rgb]{0,0,0}\hphantom{``Peppers'':} OFSE}%
\psfrag{s18}[l][l][0.75]{\color[rgb]{0,0,0}\hphantom{``Peppers'':} FSE}%
\psfrag{s19}[l][l][0.75]{\color[rgb]{0,0,0}``Baboon'':\newlength{\nbaboon}\settowidth{\nbaboon}{``Baboon'':}\hspace{-\nbaboon}\hphantom{``Peppers'':} FOFSE, $\gamma=0.2$}%
\psfrag{s20}[l][l][0.75]{\color[rgb]{0,0,0}\hphantom{``Peppers'':} OFSE}%
\psfrag{s21}[l][l][0.75]{\color[rgb]{0,0,0}\hphantom{``Peppers'':} FSE}%
\psfrag{x12}[t][t][0.75]{$0$}%
\psfrag{x13}[t][t][0.75]{$50$}%
\psfrag{x14}[t][t][0.75]{$100$}%
\psfrag{x15}[t][t][0.75]{$150$}%
\psfrag{x16}[t][t][0.75]{$200$}%
\psfrag{x17}[t][t][0.75]{$250$}%
\psfrag{x18}[t][t][0.75]{$300$}%
\psfrag{x19}[t][t][0.75]{$350$}%
\psfrag{x20}[t][t][0.75]{$400$}%
\psfrag{x21}[t][t][0.75]{$450$}%
\psfrag{x22}[t][t][0.75]{$500$}%
\psfrag{v12}[r][r][0.75]{$8$}%
\psfrag{v13}[r][r][0.75]{$10$}%
\psfrag{v14}[r][r][0.75]{$12$}%
\psfrag{v15}[r][r][0.75]{$14$}%
\psfrag{v16}[r][r][0.75]{$16$}%
\psfrag{v17}[r][r][0.75]{$18$}%
\psfrag{v18}[r][r][0.75]{$20$}%
\psfrag{v19}[r][r][0.75]{$22$}%
\psfrag{v20}[r][r][0.75]{$24$}%
\psfrag{v21}[r][r][0.75]{$26$}%
\psfrag{v22}[r][r][0.75]{$28$}%
\vspace{-4mm}\includegraphics[width=0.4\textwidth]{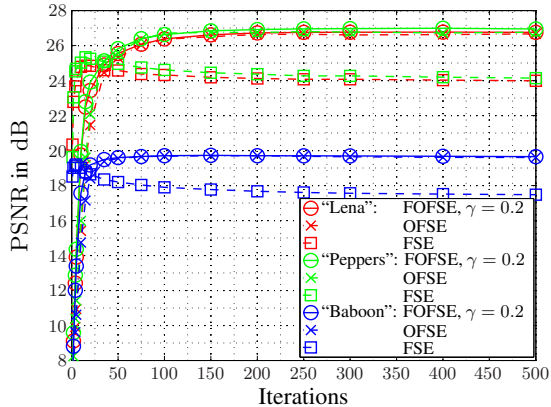}
\vspace{-0.2cm}\caption{Extrapolation quality over iterations for losses of size $16\times16$ pixels.}
	\label{fig:psnr_over_iter}
\end{figure}

In Fig. \ref{fig:psnr_over_iter} the extrapolation results are shown with respect to the number of iterations for the original frequency selective extrapolation (FSE), OFSE, and for FOFSE with $\gamma=0.2$. Although the computational complexity is reduced significantly, FOFSE is as effective as OFSE and we get an increase of up to $2 \punit{dB}$ in $\PSNR$ compared to the uncompensated extrapolation. The extrapolation properties are similar to the ones from OFSE, meaning that compared to the uncompensated extrapolation the $\PSNR$ increases a bit slower but attains a saturation level higher than the peak level of the uncompensated extrapolation.

In Table \ref{tab:odc_results} the proposed extrapolation algorithm is compared to extrapolation algorithms from Li and Orchard \cite{Li2002}, Alkachouh and Bellanger \cite{Alkachouh2000}, Sun and Kwok \cite{Sun1995}, Wang et al. \cite{Wang1993} and the frequency selective extrapolation with compensation \cite{Seiler2007} and without compensation \cite{Meisinger2004b} in terms of extrapolation quality and processing time. The mean processing time per block has been measured with MATLAB (version $7.3$) implementations on a Pentium D @ $3.2\punit{Ghz}$ with $4\punit{GB}$ RAM. The results can be split in two groups. \cite{Wang1993, Sun1995, Alkachouh2000, Meisinger2004b} provide decent extrapolation results with a small amount of processing time, whereas \cite{Li2002, Seiler2007} and the proposed algorithm perform better but at the cost of a higher computational load. Comparing the last three algorithms it becomes clear that OFSE and FOFSE perform a bit better than \cite{Li2002} regarding all test images. But by utilization of the complexity reduced compensation the real processing time can also be reduced by a factor more than $100$, reaching an acceptable level. The discrepancy between the measured processing time and the theoretic considerations in Section \ref{sec:valuation} is due to the fact, that the hypothetical memory consuming implementations are not realizable.

\begin{figure}
\centering
\includegraphics[width=0.36\textwidth]{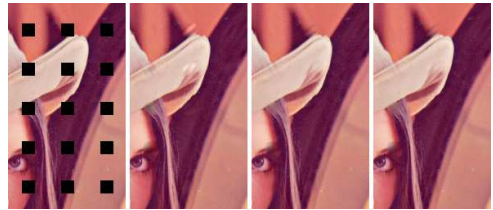}
\vspace{-0.2cm}\caption{Visual extrapolation quality for a part of test image ``Lena'' (from left to right: error pattern, FSE, OFSE, and FOFSE)}
	\label{fig:lena_combined}
\end{figure}

In order to illustrate the visual extrapolation quality, Fig. \ref{fig:lena_combined} shows a part of the erroneous image ``Lena'' concealed with FSE, OFSE, and FOFSE. Apparently, the visual extrapolation quality of OFSE and FOFSE is very similar and definitely better than with FSE.

\begin{figure}
\centering
\psfrag{s01}[t][t]{\color[rgb]{0,0,0}\setlength{\tabcolsep}{0pt}\begin{tabular}{c}Iterations\end{tabular}}%
\psfrag{s02}[b][b]{\color[rgb]{0,0,0}\setlength{\tabcolsep}{0pt}\begin{tabular}{c}$\PSNR$ in $\punit{dB}$\end{tabular}}%
\psfrag{s04}[b][b]{\color[rgb]{0,0,0}\setlength{\tabcolsep}{0pt}}
\psfrag{s06}[][]{\color[rgb]{0,0,0}\setlength{\tabcolsep}{0pt}\begin{tabular}{c} \end{tabular}}%
\psfrag{s07}[][]{\color[rgb]{0,0,0}\setlength{\tabcolsep}{0pt}\begin{tabular}{c} \end{tabular}}%
\psfrag{s08}[l][l][0.75]{\color[rgb]{0,0,0}FSE}%
\psfrag{s25}[l][l][0.75]{\color[rgb]{0,0,0}FOFSE, $\gamma = 0.1$}%
\psfrag{s26}[l][l][0.75]{\color[rgb]{0,0,0}\hphantom{FOFSE,} $\gamma = 0.3$}%
\psfrag{s27}[l][l][0.75]{\color[rgb]{0,0,0}\hphantom{FOFSE,} $\gamma = 0.5$}%
\psfrag{s28}[l][l][0.75]{\color[rgb]{0,0,0}\hphantom{FOFSE,} $\gamma = 0.7$}%
\psfrag{s29}[l][l][0.75]{\color[rgb]{0,0,0}\hphantom{FOFSE,} $\gamma = 0.9$}%
\psfrag{s30}[l][l][0.75]{\color[rgb]{0,0,0}OFSE}%
\psfrag{s31}[l][l][0.75]{\color[rgb]{0,0,0}FSE}%
\psfrag{x12}[t][t][0.75]{$0$}%
\psfrag{x13}[t][t][0.75]{$10$}%
\psfrag{x14}[t][t][0.75]{$20$}%
\psfrag{x15}[t][t][0.75]{$30$}%
\psfrag{x16}[t][t][0.75]{$40$}%
\psfrag{x17}[t][t][0.75]{$50$}%
\psfrag{x18}[t][t][0.75]{$60$}%
\psfrag{x19}[t][t][0.75]{$70$}%
\psfrag{x20}[t][t][0.75]{$80$}%
\psfrag{x21}[t][t][0.75]{$90$}%
\psfrag{x22}[t][t][0.75]{$100$}%
\psfrag{v12}[r][r][0.75]{$8$}%
\psfrag{v13}[r][r][0.75]{$12$}%
\psfrag{v14}[r][r][0.75]{$16$}%
\psfrag{v15}[r][r][0.75]{$20$}%
\psfrag{v16}[r][r][0.75]{$24$}%
\psfrag{v17}[r][r][0.75]{$28$}%
\vspace{-0.4cm}\includegraphics[width=0.4\textwidth]{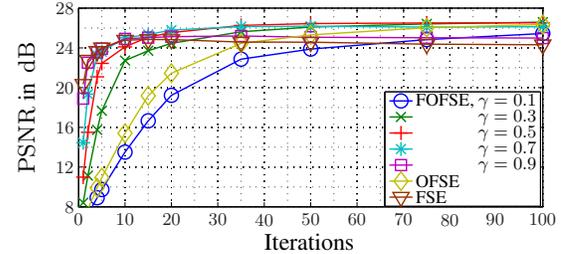}
\vspace{-0.2cm}\caption{Extrapolation quality over iterations for test image ``Lena'' and different compensation factors $\gamma$.}
	\label{fig:lena_oc}
\end{figure}

As a side effect, it is possible to tune the extrapolation between computational load and extrapolation quality by adjusting the compensation factor $\gamma$. For larger values of $\gamma$ the $\PSNR$ reaches the maximum values with less iterations although the maximum is not as high as it can be for smaller values of $\gamma$. Only for large values of $\gamma$ the degradation effect after the peak occurs as in the uncompensated extrapolation. To illustrate this circumstance, in Fig. \ref{fig:lena_oc} for test image ``Lena'' the $\PSNR$ is shown over iterations for different compensation factors $\gamma$. Thus, for every desired application it is possible to set up the extrapolation in such a way to get good extrapolation results at a fixed number of operations. 

\vspace{-2mm} \section{Conclusion} \label{sec:conclusion} \vspace{-2mm}

Our proposed algorithm is an adaption of a very powerful extra\-polation algorithm in order to cope with its high computational load. The proposed modifications significantly decrease the needed processing time without any degradation in extrapolation quality. This approach leads to very good objective and subjective results with an acceptable processing time.\vspace{-2mm} 

\end{document}